\documentclass[manuscript]{aastex}

\shorttitle{Light Brigade in Developing AR. I\hspace{-.1em}I.}
\shortauthors{Toriumi et al.}

\begin{document}


\title{Light Bridge in a Developing Active Region.\\
  I\hspace{-.1em}I. Numerical Simulation
  of Flux Emergence
  and Light Bridge Formation}


\author{Shin Toriumi$^{1}$, Mark C. M. Cheung$^{2}$, and Yukio Katsukawa$^{1}$}
\affil{$^{1}$National Astronomical Observatory of Japan, 2-21-1 Osawa, Mitaka, Tokyo 181-8588, Japan}
\email{shin.toriumi@nao.ac.jp}
\affil{$^{2}$Lockheed Martin Solar and Astrophysics Laboratory, 3251 Hanover Street, Building/252, Palo Alto, CA 94304, USA}




\begin{abstract}
Light bridges,
the bright structure
dividing umbrae
in sunspot regions,
show various activity events.
In Paper~I,
we reported on analysis
of multi-wavelength observations
of a light bridge
in a developing active region (AR)
and concluded that
the activity events are 
caused by magnetic reconnection
driven by magnetconvective evolution.
The aim of this second paper
is to investigate
the detailed magnetic and velocity structures
and the formation mechanism
of light bridges.
For this purpose,
we analyze
numerical simulation data
from a radiative magnetohydrodynamics model
of an emerging AR.
We find that
a weakly-magnetized plasma upflow
in the near-surface layers
of the convection zone
is entrained
between the emerging magnetic bundles
that appear as pores
at the solar surface.
This convective upflow
continuously transports
horizontal fields
to the surface layer
and creates
a light bridge structure.
Due to the magnetic shear
between the horizontal fields
of the bridge
and the vertical fields
of the ambient pores,
an elongated cusp-shaped current layer
is formed above the bridge,
which may be favorable
for magnetic reconnection.
The striking correspondence
between the observational results
of Paper~I
and the numerical results
of this paper
provides
a consistent physical picture
of light bridges.
The dynamic activity phenomena
occur as a natural result
of the bridge formation
and its convective nature,
which has much in common with
those of umbral dots
and penumbral filaments.
\end{abstract}





\section{Introduction
  \label{sec:introduction}}

Light bridges are
bright features
in white light
separating umbral regions
in sunspots.
It is known that
the light bridges typically
have a weaker magnetic field
compared to
the ambient umbrae
\citep{bec69,lit91,rue95}.
They are associated with
various activity phenomena
including
\ion{Ca}{2} H brightenings and ejections
\citep{lou08,shi09},
H$\alpha$ surges
\citep{roy73,asa01},
and brightenings
in {\it TRACE} 1600 {\AA} channel
\citep{ber03}.
Such phenomena
may be caused
by the interaction
between the light bridge magnetic field
and the external field
that has a canopy structure
\citep{lek97,jur06}.

The aim of this series of papers
(this paper along with \citet{tor15a},
hereafter Paper I\defcitealias{tor15a}{Paper~I})
is to reveal the nature
of the light bridges
that appear
not in the fragmentation process
of decaying sunspots
but in the assembling phase
of developing active regions (ARs)
and the resultant occurrence
of the activity phenomena.
We approach
this problem
with a comparative study
of observations
and magnetohydrodynamic (MHD) modeling.
In \citetalias{tor15a},
we analyzed NOAA AR 11974
utilizing various observational data
and investigated
the magnetic and velocity structure
of the light bridge
and its relation
with the corresponding activity events.
As a result,
we found
in the photosphere that
the light bridge
in this developing AR
has relatively weaker,
almost horizontal field,
while the surrounding pores
have stronger, almost vertical field.
As the AR continued to grow,
the pores approached
each other
and coalesced
into a single sunspot.
While it existed,
the bridge showed
a large-scale convective upflow
composed of smaller-sized cells
and horizontal divergent outflow
of a time scale of 10 -- 15 minutes.
Above the light bridge,
a chromospheric (or upper-photospheric) brightening
was intermittently and repeatedly observed
with a typical duration
of 10 -- 20 minutes.
In addition,
repeated brightenings
in the chromosphere
preceded dark surges.
These observational results
led us to propose
that magnetic reconnection
driven by the magnetoconvection
produces the activity phenomena.
More precisely,
the convective upflow 
in the light bridge
repeatedly transports
the weak horizontal fields
to the surface layer,
which then reconnects
with the strong vertical fields
of the surrounding pores,
leading to the repetitive occurrence
of the brightening
and the surge ejections.

However,
the observational study
in \citetalias{tor15a}
leaves some open questions.
For example,
the formation mechanism
of the light bridge
is unclear.
Although \citet{kat07a} found that
the light bridges
in decaying sunspots
formed
as a chain of umbral dots,
the bridge formation
in the assembling phase
of the nascent sunspots
warrants
further investigation.
Another question is
the driver
of the large-scale flow structure
inside the light bridge.
Our photospheric observation
showed that
the bridge has a broad upflow
with a flanking downflow.
However,
the vertical structure
of the large-scale convection
and its transportation
of the magnetic flux
remained unclear.
Also,
the existence
of small-scale convection cells
and the structure
of total electric current
needed further analysis.

Such questions
may be answered
by analyzing
three-dimensional (3D) MHD simulations.
In the last decade,
the progress of
realistic modeling
of the solar magnetism
using radiative MHD codes
opened a new door
to much better understanding
of various features
in the Sun.
Among others,
the {\it MURaM} code
\citep{voe05,rem09b},
which takes into account
the radiative energy transfer
and the realistic equation of state,
has been used extensively
to model a variety of magnetic structures
that are coupled
with thermal convection,
such as sunspot umbrae and umbral dots
\citep{sch06},
sunspots and penumbrae
\citep{rem09a,rem09b},
emerging flux regions
\citep{che07,che08},
and AR formations
\citep{che10,rem14}.

In Paper~II,
we provide the analysis
of the {\it MURaM} simulation
of an AR-scale
flux emergence
conducted by \citet{che10}.
In this simulation,
the
model proto-spots
are accompanied
by transient light bridges.
Although the formation and disappearance
of the light bridge
were reported in \citet{che10},
they did not pursue
a detailed investigation
of the light bridge.
We present here
an analysis
of magnetic and flow structures
of the model light bridge
in their simulations
within the context
of the observations
described in \citetalias{tor15a}.
The rest of this paper
proceeds as follows.
In Section \ref{sec:simulation},
we describe
the numerical simulation
that we use,
while in Section \ref{sec:results},
we show the analysis results.
Then,
in Section \ref{sec:discussion},
we discuss
the comparison
with observational results
and the generality
of magnetoconvection
in a strong background magnetic field.
Finally,
in Section \ref{sec:concluding},
we conclude
with a synthesis of lessons
learned from the two papers.

\section{Numerical Simulation
  \label{sec:simulation}}

In this study,
we used the numerical results
of the radiative MHD simulation
of a large-scale flux emergence
from the convection zone
by \citet{che10}.
The simulation was conducted
using the {\it MURaM} code.
In this experiment,
we set a 3D Cartesian
computational domain
of $92.16\ {\rm Mm}\times 49.152\ {\rm Mm}\times 8.192\ {\rm Mm}$,
which is spanned by
$1920\times 1024\times 256$ grid cells.
The grid spacings are
48 km for both horizontal directions
and 32 km for vertical direction.
The base of the photosphere,
which is defined
as the mean height
where the Rosseland optical depth
becomes unity
($\tau=1$),
is at $7.52\ {\rm Mm}$
above the bottom
of the domain.

The emergence of a twisted flux tube
was mimicked
by kinematically inserting
a twisted flux tube
with the shape
of a half torus
\citep{fan03,hoo09}
through the bottom boundary
into radiatively driven
near-surface convective flows.
The total magnetic flux
contained in the flux tube is
$7.6\times 10^{21}\ {\rm Mx}$,
while the field strength
averaged over the torus cross-section
is 9 kG.
After the half torus
is completely inserted
into the convection zone
(about 5.9 hours
after the start
of inserting the flux tube),
the bottom boundary
is such that
the subsurface roots
of the tube are anchored.
Magnetic field
at the top boundary
(about 670 km above
the base of the photosphere)
is matched to
a potential field,
whereas the periodic boundaries
are assumed
for both horizontal directions.

\begin{figure}
  \begin{center}
    \includegraphics[width=160mm]{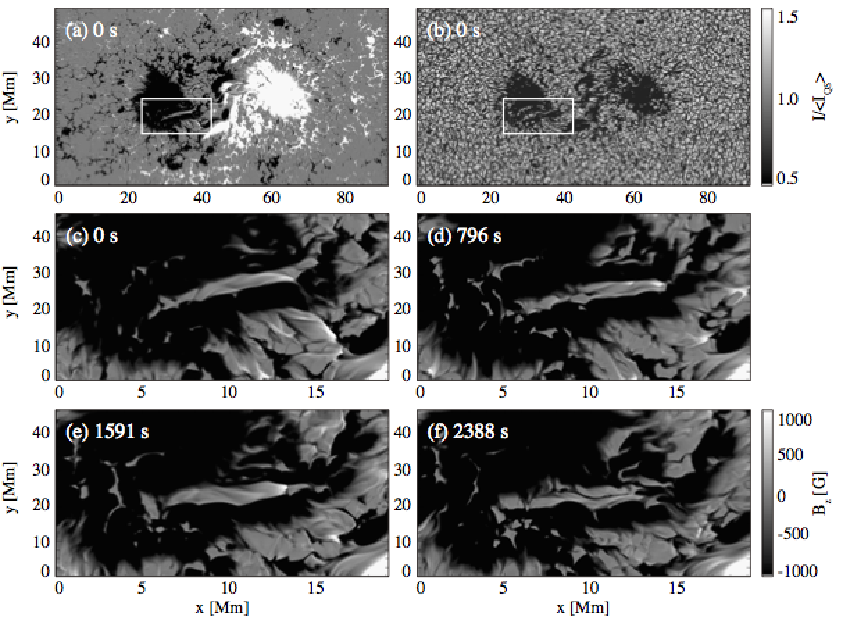}
  \end{center}
  \caption{(a) Vertical magnetic field,
    $B_{z}$,
    at $t=0\ {\rm s}$
    sampled at the base
    of the photosphere
    ($z=0\ {\rm Mm}$).
    (b) Continuum intensity map
    at the same time.
    (c--f) Sequential magnetogram
    showing the evolution
    of the light bridge structure.
    Field of view is indicated
    in panels (a) and (b)
    as a white box.    
  }
  \label{fig:general}
\end{figure}

In this particular study,
in order to obtain
simulation outputs
with a better temporal cadence,
we restarted the calculation
by using the original snapshot
as the initial condition.
The snapshot is taken
at about 19 hours
after the flux is inserted
and we define this time
as $t=0\ {\rm s}$.
Figure \ref{fig:general}(a) shows
the surface vertical field
of the entire computational domain
(i.e., $\sim 92\ {\rm Mm}\times 49\ {\rm Mm}$)
sampled at the constant height
where $<\tau>=1$
at the time $t=0\ {\rm s}$.
At this moment,
the flux tube
appears at the surface
and creates 
a pair of strong flux concentrations,
which are observed
as pores (proto-spots)
in the continuum intensity map
(Figure \ref{fig:general}(b)).
Here,
one may find that
a partial light bridge structure
is present
in the negative polarity
(see the white box
in these figures).
Figures \ref{fig:general}(c) to (f)
are the expanded view
of this region,
showing a temporal evolution
of the light bridge
for about 40 minutes.
One can see that
the light bridge has
a much weaker field
and is surrounded
by the strong negative pores.
Hereafter,
we define
3D local coordinates
$(x, y, z)$,
where $z$ is
anti-parallel with
the gravitational acceleration
and directed upward.
The origin,
$(x, y)=(0, 0)$,
is located
at bottom left
of the box in Figure \ref{fig:general}(a),
while $z=0$ is
at the base
of the photosphere
(i.e., 7.52 Mm above the bottom boundary).
We analyze the nature
of this light bridge
in detail
in Section \ref{sec:results}.

\section{Results
  \label{sec:results}}

\subsection{Magnetic and Velocity Structures
  of the Light Bridge
  \label{subsec:structure}}

\begin{figure}
  \begin{center}
    \includegraphics[width=160mm]{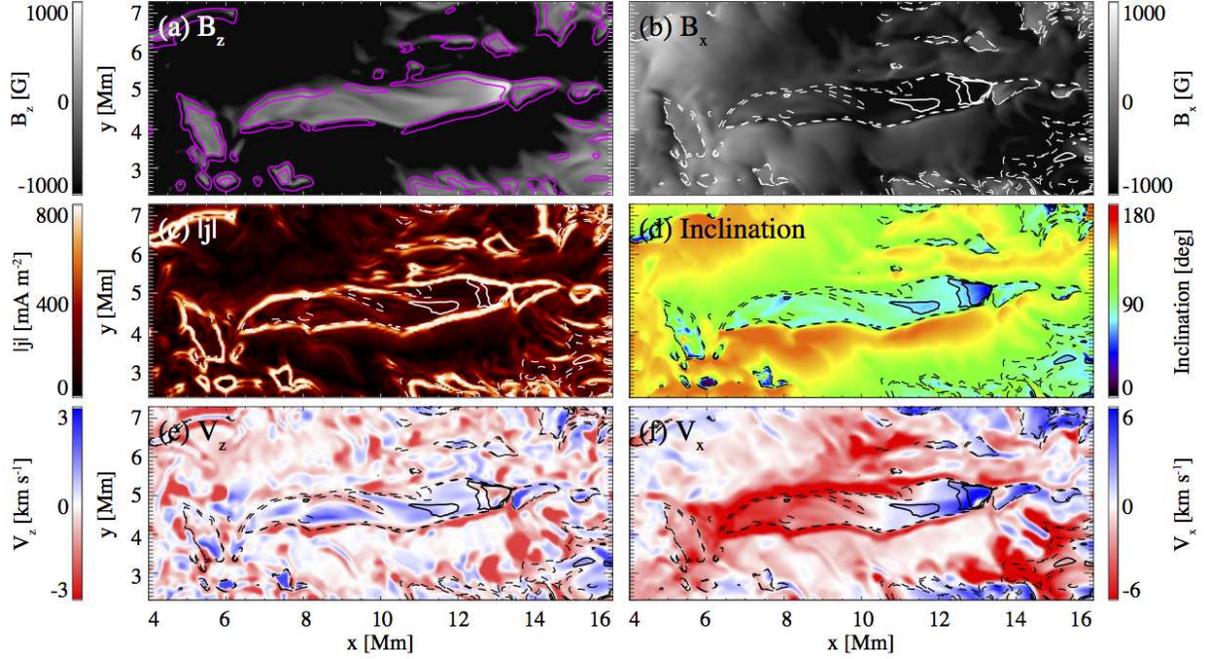}
  \end{center}
  \caption{Maps for various physical parameters
    on the $z=0\ {\rm Mm}$ plane,
    measured at $t=1591\ {\rm s}$.
    (a) Vertical field $B_{z}$,
    (b) horizontal field $B_{x}$,
    (c) total electric current density
    $|\mbox{\boldmath $j$}|$,
    (d) inclination angle of the magnetic field
    with respect to the vertical direction,
    where $0^{\circ}$ is parallel
    to  $\hat{\mbox{\boldmath $z$}}$,
    (e) vertical velocity $V_{z}$,
    where the positive value
    indicates upward,
    and (f) horizontal velocity $V_{x}$.
    In panel (a),
    purple contours
    show the current density
    of $|\mbox{\boldmath $j$}|=550\ {\rm mA\ m}^{-2}$.
    In the other panels,
    solid and dashed contours show
    $B_{z}=400\ {\rm G}$ and $200\ {\rm G}$ levels
    and $B_{z}=-200\ {\rm G}$ and $-400\ {\rm G}$ levels,
    respectively.
  }
  \label{fig:lb}
\end{figure}

Figure \ref{fig:lb} displays
the maps for various physical quantities
around the light bridge
taken at the $z=0\ {\rm Mm}$ slice
at the time $t=1591\ {\rm s}$.
The light bridge has a weak vertical field
$B_{z}$
(panel (a))
and field lines are much inclined
to the negative $x$-direction
(panels (b) and (d)).
The surrounding pores have
relatively stronger,
more vertical fields
($B_{z}<-1000\ {\rm G}$:
panels (a) and (d)).
The bridge structure
is clearly surrounded
by a wall
of the strong total electric current density
(panels (a) and (c)),
\begin{eqnarray}
  |\mbox{\boldmath $j$}|
  =\left|
    \frac{1}{\mu_{0}}
    \nabla\times\mbox{\boldmath $B$}
  \right|,
\end{eqnarray}
where $\mu_{0}$ is the magnetic permeability.
This is due to the magnetic shear
between the inclined fields
of the light bridge
and the vertical fields
of the ambient pores.
From the electric current map,
the size of the bridge
is measured to be
about $7.5\ {\rm Mm}\times 1.5\ {\rm Mm}$.
In the
vertical velocity map
($V_{z}$: panel (e)),
within the light bridge,
one may find a broad upflow region
that is divided by
narrower downflow lanes.
That is,
the light bridge contains
several convection cells inside
($\sim 2\ {\rm Mm}\times 0.5\ {\rm Mm}$).
At the edges
of the bridge,
narrow downflow lanes
are also seen.
The horizontal velocity $V_{x}$ distributions
(panel (f)) show
a divergent structure
in the light bridge:
$V_{x}$ is negative
for $x\lesssim 10.5\ {\rm Mm}$
and is positive
in the other side.
However,
at the light bridge boundary,
$V_{x}$ shows a strong negative flow
(up to $10\ {\rm km\ s}^{-1}$),
even in the range of
$x>10.5\ {\rm Mm}$.

\begin{figure}
  \begin{center}
    \includegraphics[width=160mm]{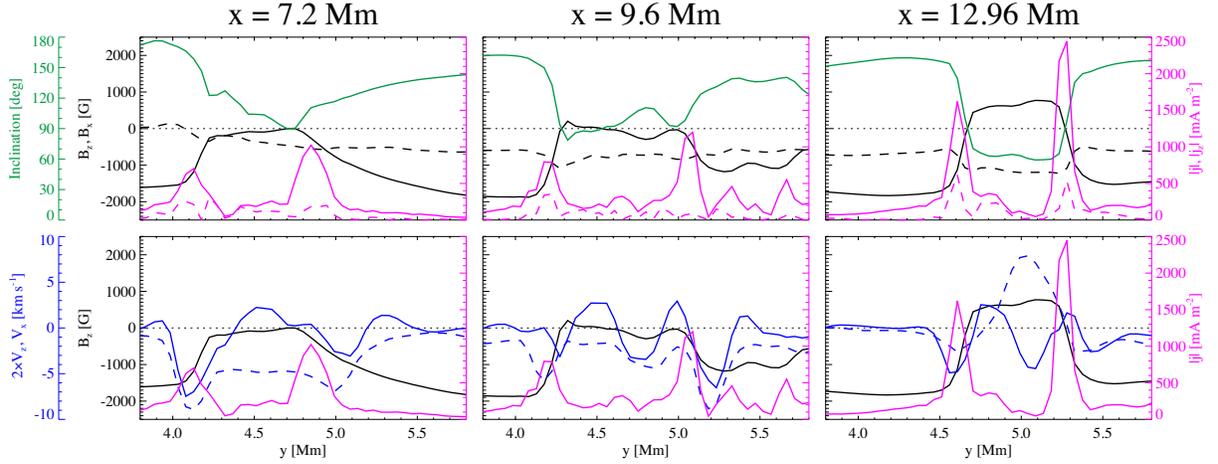}
  \end{center}
  \caption{One-dimensional ($y$-)profiles
    across the light bridge
    on the $z=0\ {\rm Mm}$ surface,
    sampled at $x=7.2\ {\rm Mm}$, 9.6 Mm, and 12.96 Mm
    at $t=1591\ {\rm s}$.
    In upper panels,
    vertical field $B_{z}$ (black solid),
    horizontal field $B_{x}$ (black dashed),
    field inclination from the vertical (green),
    total electric current density
    $|\mbox{\boldmath $j$}|$ (purple solid),
    and vertical electric current density $|j_{z}|$ (purple dashed)
    are plotted.
    In lower panels,
    vertical velocity $V_{z}$ (blue solid)
    and horizontal velocity $V_{x}$ (blue dashed)
    are plotted
    along with
    vertical field $B_{z}$ (black)
    and total current $|\mbox{\boldmath $j$}|$ (purple).
    In all panels,
    horizontal dotted lines show
    the $B_{z}=B_{x}=V_{z}=V_{x}=0$ level.
  }
  \label{fig:lb_1d}
\end{figure}

Figure \ref{fig:lb_1d} shows
1D profiles of physical parameters
sampled along the $y$-direction
at three different $x$ positions
($x=7.2\ {\rm Mm}$,
$9.6\ {\rm Mm}$, and $12.96\ {\rm Mm}$).
It is remarkable
at all three positions that
the boundary of the bridge
is clearly defined
by the total current density
$|\mbox{\boldmath $j$}|$,
which is enhanced
to $\gtrsim 500\ {\rm mA\ m}^{-2}$
because of the shear of magnetic fields
inside and outside of the bridge.
For comparison with the observations,
we also plot
the vertical component
of the current $|j_{z}|$,
which is derived
only from the horizontal fields
at a single layer,
\begin{eqnarray}
  |j_{z}|=
  \left|
    \frac{1}{\mu_{0}}
    \left(
      \frac{\partial B_{y}}{\partial x}
      - \frac{\partial B_{x}}{\partial y}
    \right)
  \right|.
\end{eqnarray}
Although $|j_{z}|$ is smaller than
$|\mbox{\boldmath $j$}|$
by a factor of a few to 10,
the peak positions
of both quantities
roughly overlap each other
(see, in particular,
the currents at $x=12.96\ {\rm Mm}$
in this figure).
In observations,
it is difficult to measure
$|\mbox{\boldmath $j$}|$
since it requires the vertical variations
of $B_{x}$ and $B_{y}$.
However,
the above result suggests that
the vertical current $|j_{z}|$
serves as rather
a reasonable indicator
of the light bridge boundary.
The value of the vertical current
at the boundary
is $|j_{z}|\gtrsim 100\ {\rm mA\ m}^{-2}$.

Note here that
the vertical current $|j_{z}|$
in this plot
is obtained
from the horizontal derivatives
of magnetic fields
measured at a constant height
in the simulation
($z=0\ {\rm Mm}$).
In the actual observation
(e.g., \citetalias{tor15a}),
however,
the magnetic fields
are measured
at a constant-$\tau$ surface,
whose geometrical height
may fluctuate
to some degree.
We discuss this effect
in Appendix \ref{sec:tau}.

At the center of the light bridge
at $x=9.6\ {\rm Mm}$,
the vertical component
of the magnetic field is very weak
($B_{z}\sim 0\ {\rm G}$)
and the field is almost parallel
to the solar surface
($B_{x}\sim -1000\ {\rm G}$;
inclination $\sim 90^{\circ}$).
The vertical velocity clearly
shows two convective upflows
($V_{z}\sim 1.5\ {\rm km\ s}^{-1}$)
that are separated
by a downflow lane
($V_{z}\sim -2\ {\rm km\ s}^{-1}$).
At this location,
the horizontal flow
is oriented
to the negative $x$-direction
($V_{x}\sim -4\ {\rm km\ s}^{-1}$).
At the edges of the bridge
(around $y=4.2\ {\rm Mm}$ and $5.2\ {\rm Mm}$),
both vertical and horizontal velocities
are negatively enhanced
($V_{z}<-3\ {\rm km\ s}^{-1}$;
$V_{x}<-8\ {\rm km\ s}^{-1}$)
with total (absolute) velocity being
up to $10\ {\rm km\ s}^{-1}$.
The similar trend is seen
in the leftmost end
of the bridge
at $x=7.2\ {\rm Mm}$.
However,
at this location,
the vertical field falls below zero
($B_{z}\lesssim 0\ {\rm G}$)
and thus the inclination
becomes larger ($>90^{\circ}$).
In contrast,
at the rightmost end
at $x=12.96\ {\rm Mm}$,
the vertical field
is intensified
($B_{z}>500\ {\rm G}$)
and thus the inclination
becomes smaller ($\sim 60^{\circ}$).
The stronger magnetic shear
around this site
corresponds to
the enhanced current density
($|\mbox{\boldmath $j$}|$ up to
almost $2500\ {\rm mA\ m}^{-2}$).
We can also find
a stronger downflow
($V_{z}<-2\ {\rm km\ s}^{-1}$)
with a positively enhanced horizontal velocity
($V_{x}\sim 8\ {\rm km\ s}^{-1}$).
Outside of the bridge
are the strong negative pores
($B_{z}\lesssim -1500\ {\rm G}$).
The field is more vertical
(inclination $>120^{\circ}$)
and the vertical motion
is suppressed
($V_{z}\sim 0\ {\rm km\ s}^{-1}$).

The cross-sectional (vertical) profiles
of the light bridge
at this time
($t=1591\ {\rm s}$)
are shown in
Figure \ref{fig:lb_slice}.
Here,
we can see that
the bridge
takes root deeper
in the convection zone
(as dictated by the model bottom boundary).
It has a large-scale upflow
with a typical velocity of
$V_{z}\sim 0.5\ {\rm km\ s}^{-1}$.
Inside this upflow,
the horizontal field $B_{x}$ is enhanced
and the vertical field $B_{z}$ is reduced,
resulting in
a larger inclination angle
($\sim 90^{\circ}$).
In this figure,
surfaces of constant optical depth ($\tau$)
are overplotted.
It is clearly seen that
the large-scale upflow lifts
the iso-$\tau$ levels
in the light bridge.
Because of the radiative cooling,
the rising material
in the upflow region
becomes denser
and drains back to the convection zone
at the edges of the bridge,
forming a narrow downflow lane.
This downflow is also seen
at the center of the bridge,
e.g., $(x, y, z)=(9.6\ {\rm Mm}, 4.7\ {\rm Mm}, -0.1\ {\rm Mm})$,
dividing the upflow region into two.
This may account for
the multi-cell structure
observed in
Figures \ref{fig:lb} and \ref{fig:lb_1d}.
Moreover,
the large-scale upflow
turns into a divergent motion
toward both $\pm x$-directions
at the top $\sim 1.5\ {\rm Mm}$ layer
in the convection zone.
This outflow
sweeps the magnetic flux
in the $\pm x$-directions
to the both ends
of the light bridge,
leading to the strong concentrations
of the vertical flux
at, e.g., around $x=13\ {\rm Mm}$.
At this location,
the $\tau=0.01$ layer
is elevated
to $z=500\ {\rm km}$.

In the $y$-$z$ slices
of Figure \ref{fig:lb_slice},
it is seen that
the horizontal fields
of the light bridge
are surrounded by
the strong downward fields
of the ambient pores.
Because of this magnetic shear,
the electric current
$|\mbox{\boldmath $j$}|$
forms a shell structure
wrapping around the light bridge.
The current layer clearly shows
a cusp structure
above the photosphere,
which is created
because the vertical fields
of surrounding pores
fan out over the light bridge
\citep[magnetic canopy:][]{lek97,jur06}.
The $y$-$z$ current maps indicate that
the enhancement of the current
observed
at the light bridge boundary
in the photosphere
of the actual Sun
\citepalias[e.g., Figure 3 of][]{tor15a}
is not due to the substantial change
of the iso-$\tau$ levels
around the bridge boundary.
In other words,
the observed strong current
is not an observational artifact
(see also Appendix \ref{sec:tau}).
In this figure
(Figure \ref{fig:lb_slice}),
it is also interesting that
the current layer
overlaps with
the fast uni-directional horizontal flow
with $V_{x}=-5\ {\rm km\ s}^{-1}$
to $-10\ {\rm km\ s}^{-1}$,
which is seen
all along the light bridge boundary,
even at the rightmost end
($x=12.96\ {\rm Mm}$).

From the above results,
the magnetic and velocity structures
of the present light bridge
can be summarized
as the ``weakly-magnetized'' intrusion
of the convective upflow
rather than the ``field-free'' intrusion.
Although relatively weak,
the light bridge still
has a horizontal field
of $B_{x}\sim -1000\ {\rm G}$
at the $z=0\ {\rm Mm}$ plane,
which is roughly
the same height
as the raised $\tau=1$ surface.
Also,
it is seen that
the light bridge
that appears above the photosphere
is only a top part
of the
deeper-rooted
convective upflow.
The cusp structure
that we observe
in the actual Sun
may be just
the tip of the iceberg.

\subsection{Field-line Structure}

Figure \ref{fig:vapor} shows
the magnetic field-line connectivity
in and around the light bridge structure.
Here we use
the VAPOR software package
developed at NCAR
\citep{cly05,cly07}
for plotting the data.
The seed points
for tracing the field lines
are selected
from the surface magnetogram,
while the color indicates
the local vertical field strength $B_{z}$.
Panel (a) shows that
all the field lines
around the light bridge
are connected to
the flux concentration
at the bottom boundary
of the domain,
which is one of the two footpoints
of the twisted emerging flux tube
inserted from the bottom boundary.
In panel (b),
we can see that
the field lines in the light bridge
(skyblue- and blue-colored fields)
are highly inclined
and are arching over
or undulating around
the surface.
These structures are created
by the local convective motions.
It is remarkable that
the field lines
of the surrounding pores
(red- and yellow-colored fields)
fan out over the horizontal fields
of the light bridge
and form a canopy structure.
The cusp-shaped current sheets
between the bridge and the pores
found in
Figures \ref{fig:lb}, \ref{fig:lb_1d}, and \ref{fig:lb_slice}
are created
by this magnetic shear.

In Figure \ref{fig:vapor}(c),
it is seen
around the photosphere
(upper part) that
the horizontal field lines
of the light bridge
(skyblue to blue)
are wrapped in the vertical fields
of the pores
(yellow to red).
Interestingly,
some of these vertical (pore) fields
are connected to
the horizontal (light bridge) fields
with downward concave parts.
This dip may be caused by the strong downflow
at the light bridge boundary
(see $y$--$z$ maps for $V_{z}$
in Figure \ref{fig:lb_slice}).
Also,
this transition of the magnetic field
from a vertical field with negative $B_{z}$
to a horizontal field with negative $B_{x}$,
which exerts
the Lorentz force (magnetic tension)
partially in the negative $x$-direction,
may be the driver
of the fast horizontal flow
in the negative $x$-direction
at the light bridge boundary
(see, e.g., Figure \ref{fig:lb}(f)).
Similar concave magnetic fields
are found
in observations.
\citet{lag14} reported
hints of
magnetic field reversals
at a bridge boundary,
interpreting that
the field lines are dragged down
by fast downflows
\citep[see also][]{sch13}.

In the deeper interior,
the field lines
from the surface
magnetic structures
converge into two magnetic bundles,
between which a weakly-magnetized region
is trapped.
This weakly-magnetized region
has a weak upflow
($V_{z}<0.5\ {\rm km\ s}^{-1}$)
and is connected
to the light bridge
in the upper layer.
This structure indicates
the close relationship
among the flux emergence,
the light bridge formation,
and the eventual sunspot formation.
Driven by the large-scale flux emergence,
the two magnetic bundles approach each other
in the convection zone,
which is observed
as the horizontal convergence
of two pores
at the solar surface
\citep{zwa85,str99}.
During this process,
in the interior,
the approaching magnetic bundles
entrain the local upflow
with weakly-magnetized plasma.
As the upflow appears
at the surface,
the light bridge structure
is created
between the two ambient pores,
which eventually form
a single sunspot.

\subsection{Flux Transportation
  in the Light Bridge
  \label{subsec:transport}}

In Section \ref{subsec:structure},
we found that
the large-scale convection pattern
of the light bridge
is a broad upflow
accompanied by horizontal diverging flow
(see Figures \ref{fig:lb}(e) and (f)).
In order to examine
whether this upflow transports
the magnetic flux
to the surface layer,
we plot in Figure \ref{fig:lb_slit_2}(a)
emergence rate of horizontal flux
that passes each depth level
(per unit time)
in the light bridge structure,
\begin{eqnarray}
  \Phi_{x}(z,t) = \int_{y_{1}}^{y_{2}}
  \left|B_{x}(z,t)\right|\, V_{z}(z,t)\, dy,
\end{eqnarray}
where $y_{1}=3.36\ {\rm Mm}$ and $y_{2}=6.24\ {\rm Mm}$.
Note that
the emergence rate
is measured
only within the upflow region
($V_{z}>0\ {\rm km\ s}^{-1}$),
i.e.,
the measured value indicates
the upwardly-transported horizontal flux.
The solid lines
in Figure \ref{fig:lb_slit_2}(a)
show temporal profiles
of the flux transport
at different depths,
ranging from $z=-3\ {\rm Mm}$ (black)
to $0\ {\rm Mm}$ (red).
We can see from this figure that
the flux is actually transported
from the deeper layer
to the surface,
although only a fraction
of the initial flux
can reach the surface.
For example,
the measured flux
at $z=0\ {\rm Mm}$
is about 50\% of that
at $z=-3\ {\rm Mm}$.
The rest of the initial flux
may return to the deeper layer
in the downflow lane
before reaching the photosphere.
Also,
a shorter-term fluctuation
is noticeable
in the profiles
of the shallower layers,
indicating that
the time scale (and the size scale)
of the convection
becomes smaller with height.
The typical period
of this fluctuation
is a few minutes.

As a result of the flux supply
by the convective upflow,
the flux
continuously
appears
at the surface layer.
Figure \ref{fig:lb_slit_2}(b) illustrates
the temporal evolution
of the surface
magnetic field $B_{z}$
along the $x$-slit
placed at the center
of the light bridge
($y=4.8\ {\rm Mm}$).
This figure clearly shows that
the separation of positive and negative polarities
repeatedly occurs
within the light bridge.
The apparent speed
of this diverging pattern
is typically $\pm 4\ {\rm km\ s}^{-1}$,
while the time scale
of this pattern
(the temporal gap
between two consecutive trails)
is a few to 10 minutes.

The V-shaped patterns
in this slit-time diagram
is due to the convective overturning
(see, e.g., Figure \ref{fig:lb}(f)),
which continuously advects
the magnetic flux
to both ends
of the light bridge.
In Figure \ref{fig:lb_slit_2}(b),
the positive (negative) polarities
are seen to drift in
the positive (negative) $x$-direction,
which reflects the fact that
the horizontal flux
in the upflow region
is oriented mostly
to the negative $x$-direction
(i.e., $B_{x}<0\ {\rm G}$).
The apparent speed
of $4\ {\rm km\ s}^{-1}$
is in good agreement with
the actual horizontal velocity
of about $4\ {\rm km\ s}^{-1}$
measured in Figure \ref{fig:lb_1d}.
Also,
the time scale of a few to 10 minutes
may reflect
the shorter-term convection
near the surface layer,
which was found
in Figure \ref{fig:lb_slit_2}(a).

\subsection{Response of the Atmosphere}

It is known that
the light bridges
produce activity phenomena
such as brightenings
and surge ejections
observed in chromospheric lines
(see Section \ref{sec:introduction}).
The possible mechanism
causing such phenomena
is magnetic reconnection
between the horizontal light bridge fields
of the vertical umbral fields.
Although we cannot reproduce
the brightenings and ejections
in the present simulation,
which only captures
the photosphere
and near-surface convection zone,
we can at least
observe the response
of upper layers
of the photosphere.

Figure \ref{fig:chrom} compares
the photospheric context
and the response in the atmosphere.
Panel (a) shows
the magnetogram $B_{z}$,
while panels (b) and (c)
are the total current density
$|\mbox{\boldmath $j$}|$
averaged over the height range
$160\ {\rm km}\le z \le 480\ {\rm km}$.
Panel (b) shows the smoothed current
and panel (c) is the original image.
Here,
the current
in panel (b)
exhibits
a filamentary structure
above the central line
of the light bridge
and is more enhanced
in the right half
($x>10.5\ {\rm Mm}$)
at this time.
In the online movie,
this filamentary structure
evolves dynamically
in association with
the convective motion
in the photosphere.
The filamentary structure
is seen simply because
the two current layers
around the light bridge
are closer
in the higher altitudes
due to the cusp shape
(panel (c),
see also Figure \ref{fig:lb_slice}).
Since such a strong current
is likely a place for heating,
an elongated bright structure
may be observed
in upper atmosphere
above the light bridge
(say, in the chromosphere).

Magnetic reconnection,
which may be observed
as a sudden enhancement
of the intensity,
preferentially occurs
at the places
with a strong magnetic shear.
In the present light bridge,
the magnetic field
is oriented mostly
to the negative $x$-direction
nd, thus,
the right half
of the bridge shows
the positive polarities
(Figure \ref{fig:lb_slit_2}(b)).
Therefore,
the magnetic shear
is stronger
in the right half,
which is reflected
in the stronger currents
in Figure \ref{fig:chrom}(b).
\citet{che12} found
that the Hall effect
in the weakly ionized photosphere
is strongest
at the cusp
of the bridge,
which may suggest that
the cusp could be a preferred location
for fast magnetic reconnection
\citep{bis95}.
In such a case,
we conjecture that
the intensity enhancement
(heating of local plasma)
and the surge ejection
(cool plasma outflow
launched from the reconnection site)
are observed
in the upper atmosphere.
Since the horizontal field
is supplied
by the repetitive convection upflow
as we saw
in Section \ref{subsec:transport},
the reconnection
and the resultant
intensity enhancement
and surge ejection
may occur repeatedly and intermittently
in synchronization with
the convection.
In other words,
the atmospheric activity
is driven by
the magnetoconvective dynamics
in the solar interior.

\subsection{Summary of the Numerical Results}

In this section,
we analyzed the numerical data
of the convective flux emergence simulation
focusing on the light bridge structure
that appeared
in one of the bipolar flux concentrations.
The light bridge had a size of
$\sim 7.5\ {\rm Mm}\times 1.5\ {\rm Mm}$
and was sandwiched
between the strong pores
of the negative polarity.
The results obtained
through a series of detailed analysis
are summarized as follows:
\begin{itemize}
\item In the photosphere,
the light bridge had
a relatively weak field,
which was highly inclined and almost horizontal
to the solar surface.
In the bridge center,
the vertical component
of the field strength
was $B_{z}\sim 0\ {\rm G}$
with an inclination of $\sim 90^{\circ}$,
oriented to the negative $x$-direction
(i.e., $B_{x}<0\ {\rm G}$).
In contrast,
the surrounding pores
had a strong negative polarity
($B_{z}\lesssim -1500\ {\rm G}$)
and the field lines
were more vertical.
The magnetic shear
between the horizontal fields
of the light bridge
and the vertical fields
of the ambient pores
formed a sharp current layer
at the edges
of the light bridge
($|\mbox{\boldmath $j$}|\gtrsim 500\ {\rm mA\ m}^{-2}$;
$|j_{z}|\gtrsim 100\ {\rm mA\ m}^{-2}$).
The bridge structure
showed a broader upflow
($V_{z}\sim 2\ {\rm km\ s}^{-1}$)
in its center,
which was divided by downflow lanes
into smaller convective cells
($\sim 2\ {\rm Mm}\times 0.5\ {\rm Mm}$).
This broad upflow
turned into the diverging, bi-directional motion
of $V_{x}\sim\pm 4\ {\rm km\ s}^{-1}$.
The light bridge also had
a narrow downflow lane
at its edge
($V_{z}<-3\ {\rm km\ s}^{-1}$).
Along this edge,
a rapid uni-directional flow
of $V_{x}<-8\ {\rm km\ s}^{-1}$
was also observed.

\item The light bridge
took root
deeper down in
the convection zone
and had a large-scale convective upflow
($V_{z}\sim 0.5\ {\rm km\ s}^{-1}$),
which transports the weak horizontal fields
to the solar surface
(the ``weakly-magnetized'' intrusion).
Surfaces of constant optical depth
(iso-$\tau$ levels)
were lifted
by this upflow.
The ascending plasma inside
was radiatively cooled
and became denser,
turning back into the convection zone
at the edges of the light bridge.
The cross-sectional slices
of the bridge
showed a cusp structure,
created due to the magnetic canopy
of the surrounding pore fields.

\item The magnetic field lines
in the light bridge
were highly inclined
and had arched or serpentine structures,
reflecting the local convection patterns.
The horizontal fields were surrounded
by the vertical fields
of the neighboring pores.
We found that
some of the surrounding vertical fields
were connected to the horizontal light bridge fields
with a slight concave dip
at the light bridge boundary.
The dip may be created
by the strong downflows
at the boundary.
Also,
by exerting the magnetic tension force,
this dip structure
may drive
the rapid uni-directional horizontal flow.
With depth,
the field lines
in and around the light bridge
converged into two flux bundles
at the bottom boundary,
between which a weakly-magnetized upflow
was trapped.
This means that,
in the process
of the merging of pores
and the formation of a sunspot,
the entrained upflow
between the flux bundles (pores)
appeared at the surface
and formed the light bridge
with supplying horizontal fields.

\item We analyzed
the transport
of the horizontal fields
by the light bridge upflow
and found that
only a fraction of the initial flux
in the deeper layer
reached the surface layer.
The rest of the flux 
may return deeper down
before reaching the surface.
For the flux evolutions
of the shallower layers,
a short-term fluctuation was noticeable
compared to those of the deeper layers.
This is possibly due to
the repeated appearance
of shorter-lived, smaller-scale convection patterns
near the surface.
The upwardly transported horizontal fields
were then guided
by the lateral diverging flow.
As a result of the flux advection,
the vertical magnetic fields showed
repeated separations
of positive and negative polarities
with an apparent speed
of $V_{x}=\pm 4\ {\rm km\ s}^{-1}$
and a time scale of 10 minutes.
The positive (negative) footpoints
moved to the positive (negative) $x$-direction,
which agreed with the fact that
the horizontal fields were oriented
to the negative $x$-direction.

\item Magnetic reconnection
between the transported horizontal fields
in the light bridge
and the pre-existing vertical fields
of the umbral surroundings
may be responsible
for various dynamic phenomena
observed in the light bridge.
We found that
the electric current
of several 100 km
above the light bridge
is observed as
a filamentary structure
due to the cusp-shaped current sheet.
The current was enhanced
around the location
where the magnetic shear is stronger.
Such a place is preferable
for the magnetic reconnection,
which may be observed
as a sudden intensity enhancement
of the chromospheric lines.
Since the magnetic flux
is transported by the convective upflow
in the light bridge,
the brightening and the associated surge ejection
may take place
repeatedly and intermittently
following the convective motion.
\end{itemize}

\section{Discussion
  \label{sec:discussion}}

\subsection{Comparison with Observations}

Figure \ref{fig:comparison} compares
the light bridge
observed in an actual AR
in \citetalias{tor15a}
and that modeled
in the present MHD simulation.
The observation data
were obtained by
the spectropolarimeter
of {\it Hinode}/SOT
(\citealp{kos07,tsu08,lit13};
see Section 2.1 of \citetalias{tor15a}
for details).
One may see that
the structures of the bridges
are remarkably consistent
with each other:
compare also
Figures 3(g) and 4
of \citetalias{tor15a}
and Figures \ref{fig:lb_1d}
and \ref{fig:lb_slit_2}(b)
of this paper.
Table \ref{tab:comparison} summarizes
various photospheric parameters.
Except for the parameters
representing the overall evolution
(size and lifetime of the light bridge),
the properties
of the two cases
agree well with each other.
It should be noted here that
the geometrical heights
at which the photospheric parameters
are measured
are not necessarily the same.
The parameters
in the observation
are measured
using photospheric \ion{Fe}{1} lines,
which may geometrically fluctuate
to some degree,
while the parameters
in the simulation
are measured
at a single horizontal plane
at $z=0\ {\rm Mm}$.
Also,
in Table \ref{tab:comparison},
the vertical velocity $V_{z}$
in the observation
is actually
a Doppler (line-of-sight) velocity.
Nevertheless,
the two cases show
a striking correspondence,
indicating that
the numerical simulation
captures some important physics
of light bridges.
The clear consistency
of the physical quantities
also indicates that
we may be able to use
the numerical results
to estimate the physical states
of the magnetic fields
below and above the photosphere,
which are difficult
to directly observe.

Regarding the magnetic configuration
above the photosphere,
\citet{lek97} and \citet{jur06}
proposed a canopy structure
above the light bridge.
Our simulation results
support their interpretation
(Figures \ref{fig:lb_slice}
and \ref{fig:vapor}).
\citet{shi09} reported
strong electric current
($|j_{z}|>100\ {\rm mA\ m}^{-2}$)
along a light bridge
in the photosphere
and long-lasting chromospheric jets
intermittently and recurrently
emanating from the bridge.
They interpreted
this enhanced current
as a current-carrying, highly twisted flux tube
that is trapped
below a magnetic cusp,
which reconnects with
the surrounding vertical umbral field
to trigger the chromospheric ejections
\citep{nis12}.
\citet{shi09} observed
another current enhancement,
which they interpreted
as a current sheet
between the horizontal twisted flux tube
and the vertical umbral field.
In the present simulation,
we also found two lanes
of electric current
of almost the same magnitude
($|j_{z}|\gtrsim 100\ {\rm mA\ m}^{-2}$).
Contrary to the flux tube model
by \citet{shi09},
the present current layers
are formed
due simply to
the magnetic shear
between the vertical pore field
and the horizontal bridge field,
which is transported
by a large-scale convective upflow.
Chromospheric brightenings
above the light bridge
may be observed as
a tube-like, filamentary structure
as in Figure \ref{fig:chrom}(b).
However,
we should take care
that such a structure
does not necessarily
show the existence
of the flux tube,
as is indicated
in Figure \ref{fig:chrom}(c).

\subsection{Magnetoconvection
  in a Strong Background Magnetic Field}

From the observational and numerical results
of this series of papers,
we found that
the light bridges
have a convective upflow
into a strong background magnetic field
(``weakly-magnetized'' intrusion).
We suggested that
the magnetic shear between
the horizontal fields
transported by this upflow
and the vertical fields
of ambient pores
is essential
for various activity events
around the bridges.
As was discussed earlier
by \citet{spr06} and \citet{rim08},
similar magnetic and velocity structures
are known to exist
in the Sun,
especially in sunspot regions.

One such example
is umbral dots,
the transient bright points
in umbrae of sunspots.
It is known that
the umbral dots
have upflows with weaker field
\citep[see][and references therein]{bor11}.
\citet{par79} and \citet{cho86}
suggested that
the umbral dots
are the ``field-free'' intrusions
of hot plasma
into gappy umbral magnetic field.
\citet{sch06} reported that,
in their magnetoconvection simulation,
bright, narrow upflow plumes
(i.e. umbral dots)
with adjacent downflows
are produced
in a strong vertical magnetic field.
They found that
the plumes have
a cusp structure
and are almost field-free.

Another example is
penumbral filaments,
the radially extended thin structures
in sunspot penumbra.
Observationally,
the magnetic fields
in the penumbra
have two distinct inclinations
interlaced with each other,
which is referred to as
``uncombed penumbra''
\citep{sol93}
or ``interlocking comb structure''
\citep{tho92}.
In the penumbra,
radial outflow
called Evershed flow
is observed
\citep{eve09}.
Moreover,
\citet{kat07b} found
small-scale jetlike features
(penumbral microjets),
which are possibly caused by
magnetic reconnection
between the interlocking fields.
In fact,
the magnetoconvective pattern
of the simulated penumbra
is to a large extent
similar to
that of the light bridges
in our present study
\citep{rem09a,rem09b,rem11,rem12}.
Also,
convective overturning motions
similar to our cases
have been found
in the observations
\citep[see, e.g.,][]{zak08}.

The similarities and consistencies
among the umbral dots,
light bridges,
and penumbral filaments,
which are summarized
in Table \ref{tab:magnetoconvection},
point to the generality
of the magnetoconvection
in a strong background field.
Moreover,
such a convection may produce
variety of activity events
through magnetic reconnection.
Therefore,
we can conclude that
the magnetoconvection
is not only a common physical phenomenon
associated with the aforementioned features,
but is also the essential driver
of dynamic activity
in sunspot regions.

\section{Concluding Remarks
  \label{sec:concluding}}

In this series of papers,
we studied the nature
of light bridges
in newly developing ARs.
From these studies,
we present
a consistent physical picture
of light bridges
based on
both observations and numerical simulations.

The formation
of a light bridge
in an emerging flux region
is shown in
Figure \ref{fig:illust}(a).
The magnetic fields emerge
through the convection zone
in the form of
split multiple
flux bundles,
which are observed
as fragment polarities (pores)
in the photosphere
\citep{zwa85}.
As the flux bundles emerge,
weakly-magnetized local plasma
with upflow
is entrained
between the bundles
and appears
as a light bridge
at the visible surface.
The pores of the same polarities
merge together
and, eventually,
sunspots are formed.

As shown in Figure \ref{fig:illust}(b),
the bridge has a deep convective upflow
in its center,
which carries
horizontal magnetic fields
to the surface layer.
Due to the radiative cooling,
the ascending plasma
loses buoyancy and
sinks down
at the narrow downflow lanes.
At the bridge boundaries,
the magnetic shear
between the horizontal fields
of the bridge
and the vertical fields
of the ambient pores
forms a strong electric current layer.
Some external vertical fields
are connected to the internal horizontal field
at the bridge boundary
with concave dips,
which are caused by strong downflows.
In the upper atmosphere
above the light bridge,
the external vertical fields
fan out
and form canopy structures.

Various activity phenomena
come about
as a natural result
of the bridge formation.
In the cusp-shaped current layer
formed above the light bridge,
magnetic reconnection
takes place.
The heating
of local plasma
is observed as
intensity enhancements
of chromospheric (or upper-photospheric) lines,
while the reconnection outflow
of cool, dense plasma
is observed
as dark surges
ejected into the coronal heights.
Since the horizontal flux
is continuously provided
by convection,
the activity phenomena
last longer
with a periodic and intermittent nature.

The large-scale velocity structure
is depicted in Figure \ref{fig:illust}(c).
Along with the convection
within the cross-section,
the upflow turns
also in the direction
along the length
of the bridge.
The diverging (bi-directional) outflow
transports the magnetic flux
to both ends of the bridge.
Interestingly,
rapid uni-directional horizontal flows
are observed,
which may be driven
by the bent field lines
connecting external vertical fields
and internal horizontal fields.
In the shallower layers,
smaller-scale, shorter-lived
convection cells are superposed.
As \citet{che12} pointed out,
the Hall effect
may generate the velocity
and (as a result of advection)
the magnetic field
in the direction
of bridge axis.

The above magnetoconvection scenario
provides a unified perspective
to wider variety
of features
in sunspot regions
including umbral dots
and penumbral filaments.
All these features show
a magnetoconvection
in a strong background field.
In fact,
the magnetic and velocity structures
of the light bridges,
as well as the resultant activity events
in the upper atmosphere,
are remarkably similar
to those of penumbral filaments
\citep{rem12,kat07b}.

Although we have conducted
a thorough investigation,
we still have some remaining problems
due to the limitations
of the current analysis.
Observationally,
the magnetic structure
at and around the magnetic reconnection
that produces the activity phenomena
is not determined.
So far,
{\it Hinode} data
used in \citetalias{tor15a}
provides only a vector magnetogram
at a single photospheric plane,
which is lower than
the expected reconnection site
in the chromosphere
or in the upper photosphere.
Future missions
may provide the detailed field information
not only in the photosphere
but also in the upper atmosphere.
Another issue
may be the dynamics
of the weakly ionized plasma.
In such altitudes,
partial ionization effects
such as the Hall effect
and the ambipolar diffusion,
which are not taken into account
in the present simulation,
may play important roles
in the light bridge formation
and the resultant reconnection
\citep{lea14,mar15}.
Future investigations
with highly resolved simulations
including these effects
will further reveal
the nature of the light bridge.

\acknowledgments

The authors thank
the anonymous referee
for improving the manuscript.
{\it Hinode} is a Japanese mission
developed and launched by ISAS/JAXA,
with NAOJ as domestic partner
and NASA and STFC (UK)
as international partners.
It is operated by these agencies
in co-operation with ESA and NSC (Norway).
This work was supported by
JSPS KAKENHI Grant Number 26887046
(PI: S. Toriumi)
and 25220703
(PI: S. Tsuneta). 
MCMC acknowledges support
by NASA grant NNX14AI14G
(Heliophysics Grand Challenges Research).

\appendix

\section{Effect of Optical Depths
  \label{sec:tau}}

Figure \ref{fig:lb_taudepend} shows
various physical parameters
measured at a constant geometrical height
($z=0\ {\rm Mm}$)
and three different optical depths
($\tau=1$, $0.1$, and $0.01$).
The general structure
looks consistent
with each other.
According to \citet{bor14},
in deriving magnetic and velocity parameters,
Milne-Eddington inversion codes
generally sample shallower atmospheric layers
(e.g., $\log{\tau}\sim-1.4$ for field strength,
$\log{\tau}\sim-1.7$ for inclination,
and $\log{\tau}\sim-1.0$ for line-of-sight velocity).
Therefore,
the {\it Hinode} observations
in Figure \ref{fig:comparison}
may be comparable to
values at $\tau=0.1$ or $0.01$
in Figure \ref{fig:lb_taudepend}.

\begin{figure}
  \begin{center}
    \includegraphics[width=155mm]{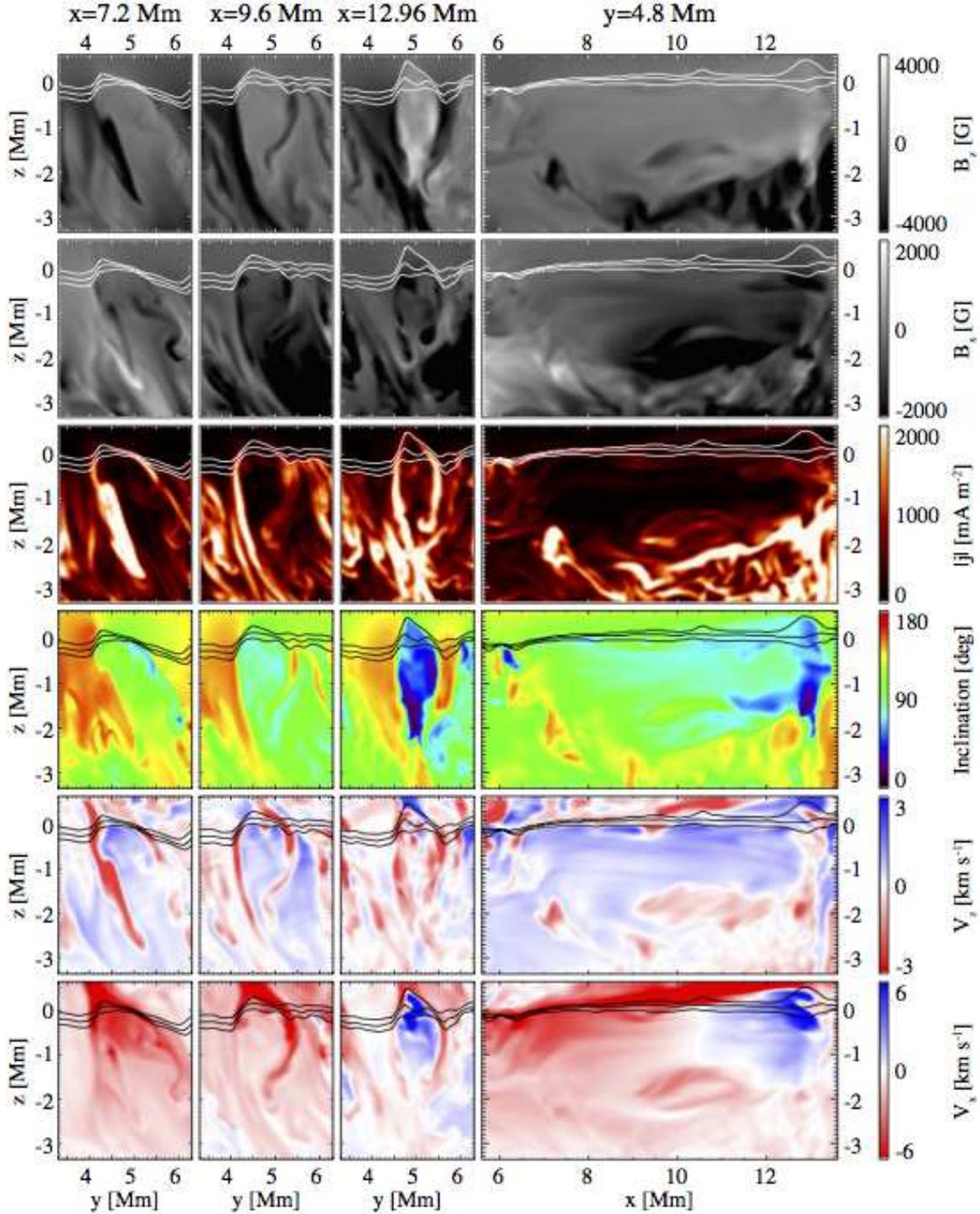}
  \end{center}
  \caption{Cross-sectional vertical slices
    at $x=7.2\ {\rm Mm}$,
    $9.6\ {\rm Mm}$,
    $12.96\ {\rm Mm}$,
    and $y=4.8\ {\rm Mm}$
    at the time $t=1591\ {\rm s}$.
    The panels show
    from top to bottom
    vertical field $B_{z}$,
    horizontal field $B_{x}$,
    total current density $|\mbox{\boldmath $j$}|$,
    inclination,
    vertical velocity $V_{z}$,
    and horizontal velocity $V_{x}$.
    The three solid lines
    in each panel indicate
    the $\tau=1$, $0.1$, and $0.01$ levels.
  }
  \label{fig:lb_slice}
\end{figure}

\begin{figure}
  \begin{center}
    \includegraphics[width=70mm]{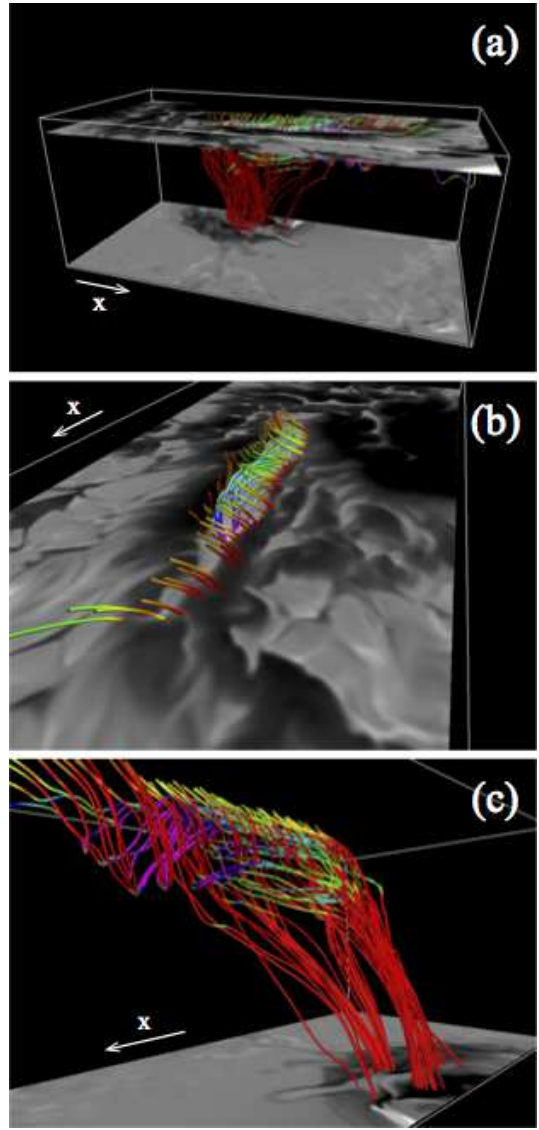}
  \end{center}
  \caption{3D visualization
    of magnetic field lines
    at the time $t=1591\ {\rm s}$.
    Seed points for the field lines
    are chosen in and around
    the light bridge structure
    observed at $z=0\ {\rm Mm}$.
    The color of the field lines represents
    the local vertical field strength $B_{z}$
    ranging from $-1000\ {\rm G}$ (red)
    to $1000\ {\rm G}$ (purple).
    The upper horizontal plane
    is a magnetogram
    at the mean $\tau=1$ surface
    (i.e., $z=0\ {\rm Mm}$).
    The grayscale saturates
    at $-2000\ {\rm G}$ (black)
    and $2000\ {\rm G}$ (white).
    The lower plane
    is the same
    but near the bottom boundary
    ($z=-7.52\ {\rm Mm}$),
    saturating at $-20,000\ {\rm G}$ (black)
    and $20,000\ {\rm G}$ (white).
    In panel (c),
    the surface magnetogram
    is not shown.
  }
  \label{fig:vapor}
\end{figure}

\begin{figure}
  \begin{center}
    \includegraphics[width=100mm]{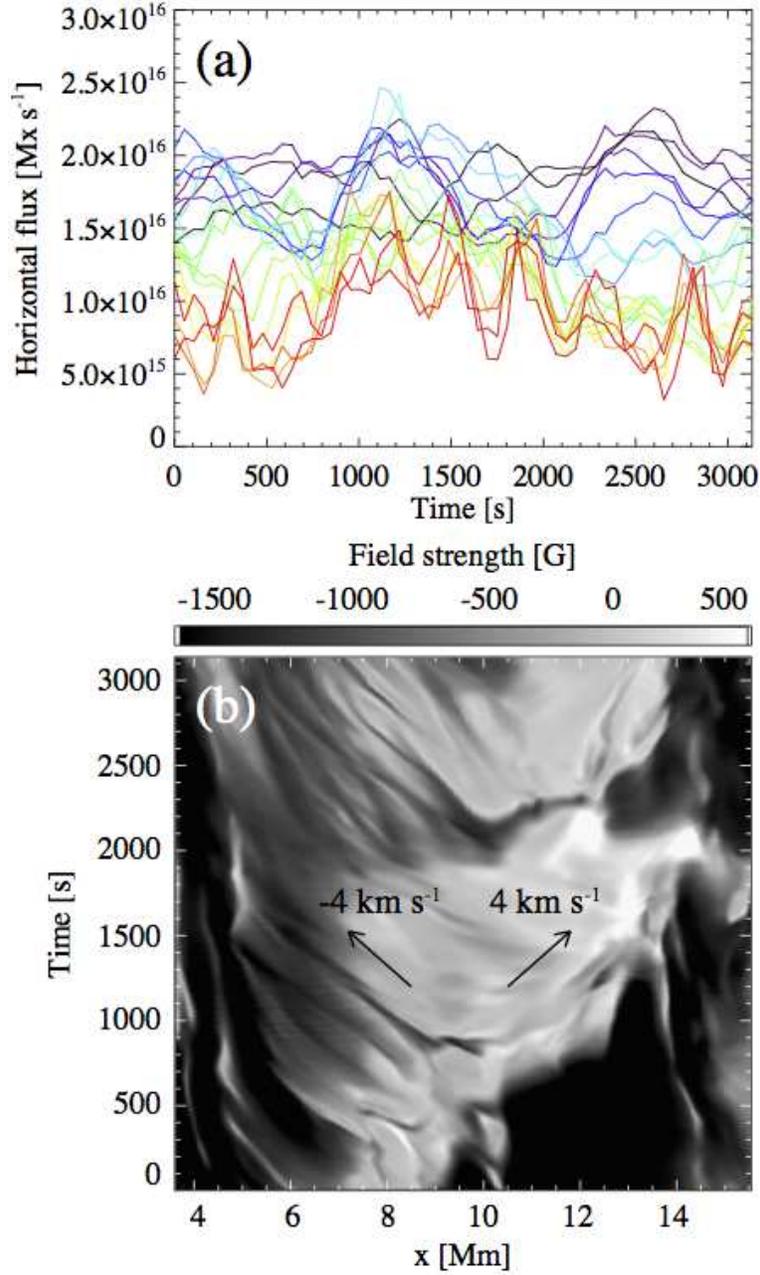}
  \end{center}
  \caption{(a) Temporal evolution
    of the upwardly-transported
    unsigned horizontal flux
    (per unit time)
    $\Phi_{x}(z,t)$
    (see text for details).
    Colors indicate the depths
    at which the flux is measured,
    ranging from $z=-3\ {\rm Mm}$ (black)
    to $z=0\ {\rm Mm}$ (red).
    (b) Temporal evolution
    of the vertical magnetic field $B_{z}$
    at $z=0\ {\rm Mm}$
    along the slit.
    The slit is placed at $y=4.8\ {\rm Mm}$,
    while arrows indicate
    the two different velocities.}
  \label{fig:lb_slit_2}
\end{figure}

\begin{figure}
  \begin{center}
    \includegraphics[width=120mm]{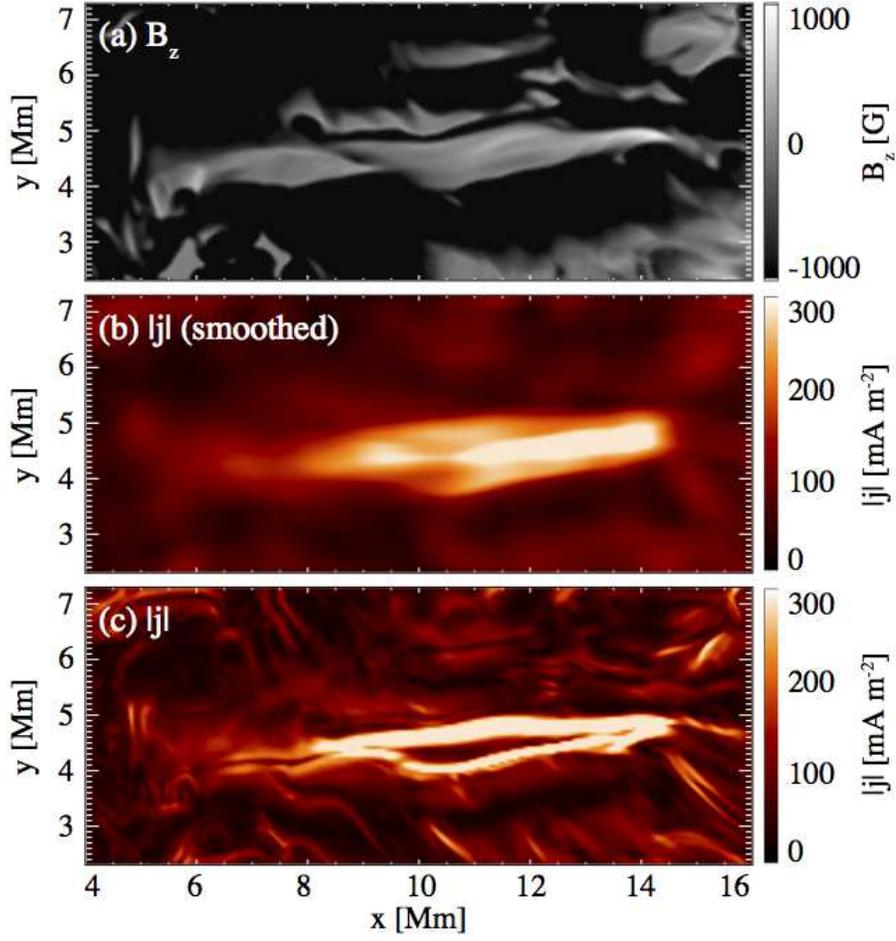}
  \end{center}
  \caption{(a) Vertical field strength $B_{z}$
    at $z=0\ {\rm km}$
    at $t=1964\ {\rm s}$.
    (b) Total electric current density
    $|\mbox{\boldmath $j$}|$
    averaged over $160\ {\rm km}\leq z\leq 480\ {\rm km}$
    and smoothed for horizontal directions.
    (c) Same as (b) but without smoothing.}
  \label{fig:chrom}
\end{figure}

\begin{figure}
  \begin{center}
    \includegraphics[width=155mm]{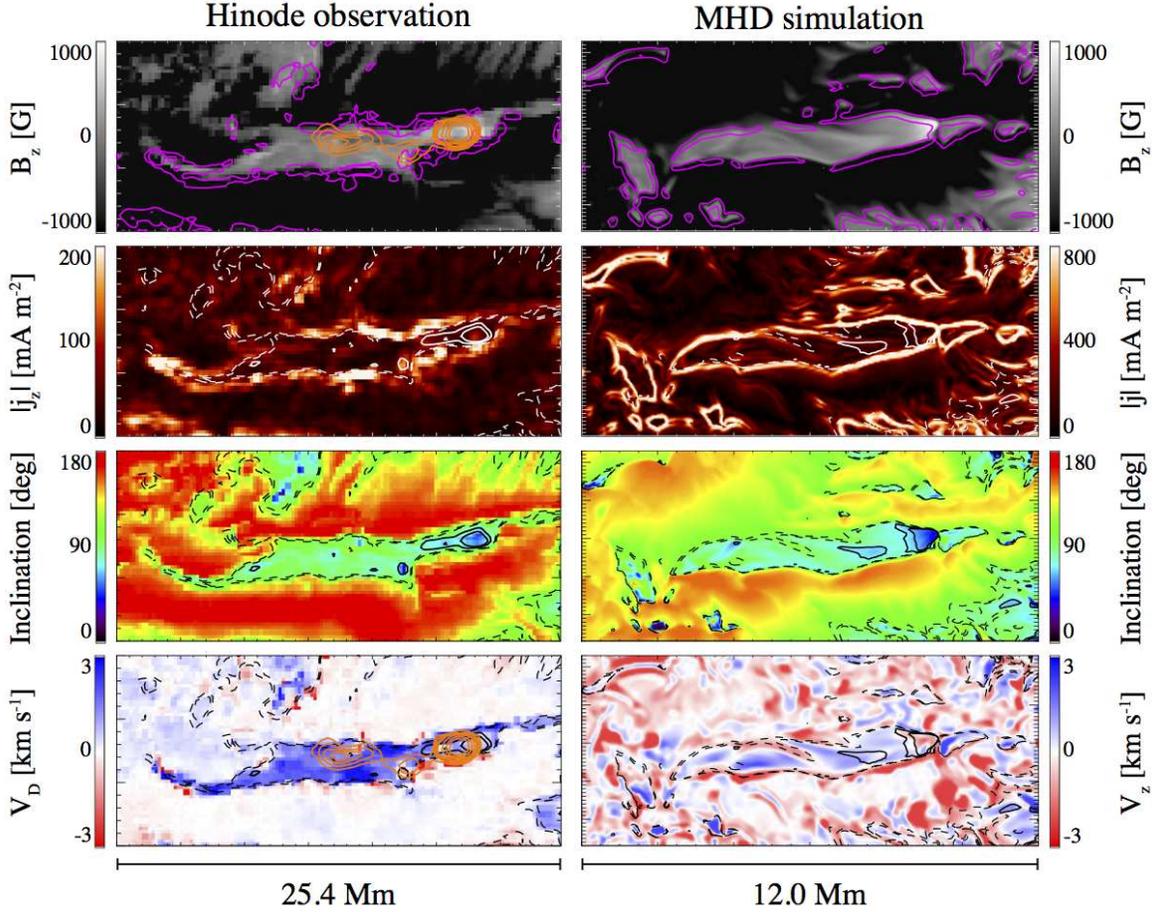}
  \end{center}
  \caption{Comparison
    of the observed and simulated
    light bridges.
    (Left) Light bridge structure
    in NOAA AR 11974
    observed by {\it Hinode}/SOT/SP.
    Figures are reproduced
    from Figure 3 of \citetalias{tor15a}.
    From top to bottom,
    vertical field $B_{z}$,
    vertical current $|j_{z}|$,
    field inclination
    with respect to the local vertical,
    and Doppler velocity $V_{\rm D}$
    are shown.
    Purple and orange contours
    indicate vertical currents
    and chromospheric intensity levels,
    respectively.
    Solid and dashed lines are
    $B_{z}=400\ {\rm G}$ and $200\ {\rm G}$ levels
    and $B_{z}=-200\ {\rm G}$ and $-400\ {\rm G}$ levels,
    respectively.
    See \citetalias{tor15a}
    for details.
    (Right) Results
    of the MHD simulation
    carried out in this paper.
    Figures are reproduced
    from Figure \ref{fig:lb}.
    From top to bottom,
    vertical field $B_{z}$,
    total current $|\mbox{\boldmath $j$}|$,
    field inclination,
    and vertical velocity $V_{z}$
    are shown.
    The physical values
    are sampled
    at $z=0\ {\rm Mm}$.
    Purple contours
    indicate the total current,
    while solid and dashed contours
    are the same as
    those in observation plots.
    See captions for Figure \ref{fig:lb}
    for details.}
  \label{fig:comparison}
\end{figure}

\begin{figure}
  \begin{center}
    \includegraphics[width=60mm]{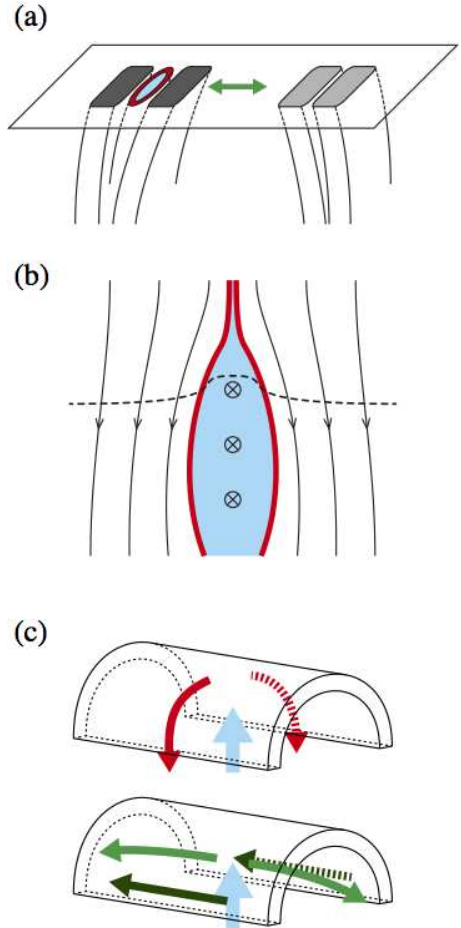}
  \end{center}
  \caption{
    {\footnotesize
    (a) Light bridge formation
    in an emerging flux region.
    Darker and lighter rounded rectangles
    are the pores,
    from which the magnetic fields
    are extended downward
    into the convection zone.
    On the left side,
    a light bridge structure
    is sandwiched
    between the pores.
    The brigade has
    a broad upflow region (blue)
    with a narrow downflow lane (red).
    Flux emergence shows
    a diverging motion (green arrow),
    which results in the coalescence
    of pores of the same polarities.
    (b) Cross-sectional view
    of the light bridge
    and the surrounding pores.
    The light bridge has an upflow (blue),
    which transports a horizontal magnetic flux
    to the surface layers.
    The iso-$\tau$ levels are elevated
    by this upflow:
    dashed line indicates
    the $\tau=1$ level.
    Narrow downflow lanes (red)
    are formed
    at the bridge boundary.
    Outside of the bridge
    are the vertical magnetic fields
    of the surrounding pores.
    (c) Illustration showing
    the convection patterns
    within the cusp structure.
    In the light bridge,
    the plasma shows an upflow (blue),
    which diverges
    and turns into downflows (red)
    at the bridge boundary.
    The upflow also diverges
    to the direction
    of the light bridge,
    showing a bi-directional flow (light green).
    At the bridge boundary,
    a fast uni-directional flow (dark green)
    is also formed.
    }
  }
  \label{fig:illust}
\end{figure}

\begin{figure}
  \begin{center}
    \includegraphics[width=200mm,angle=90.]{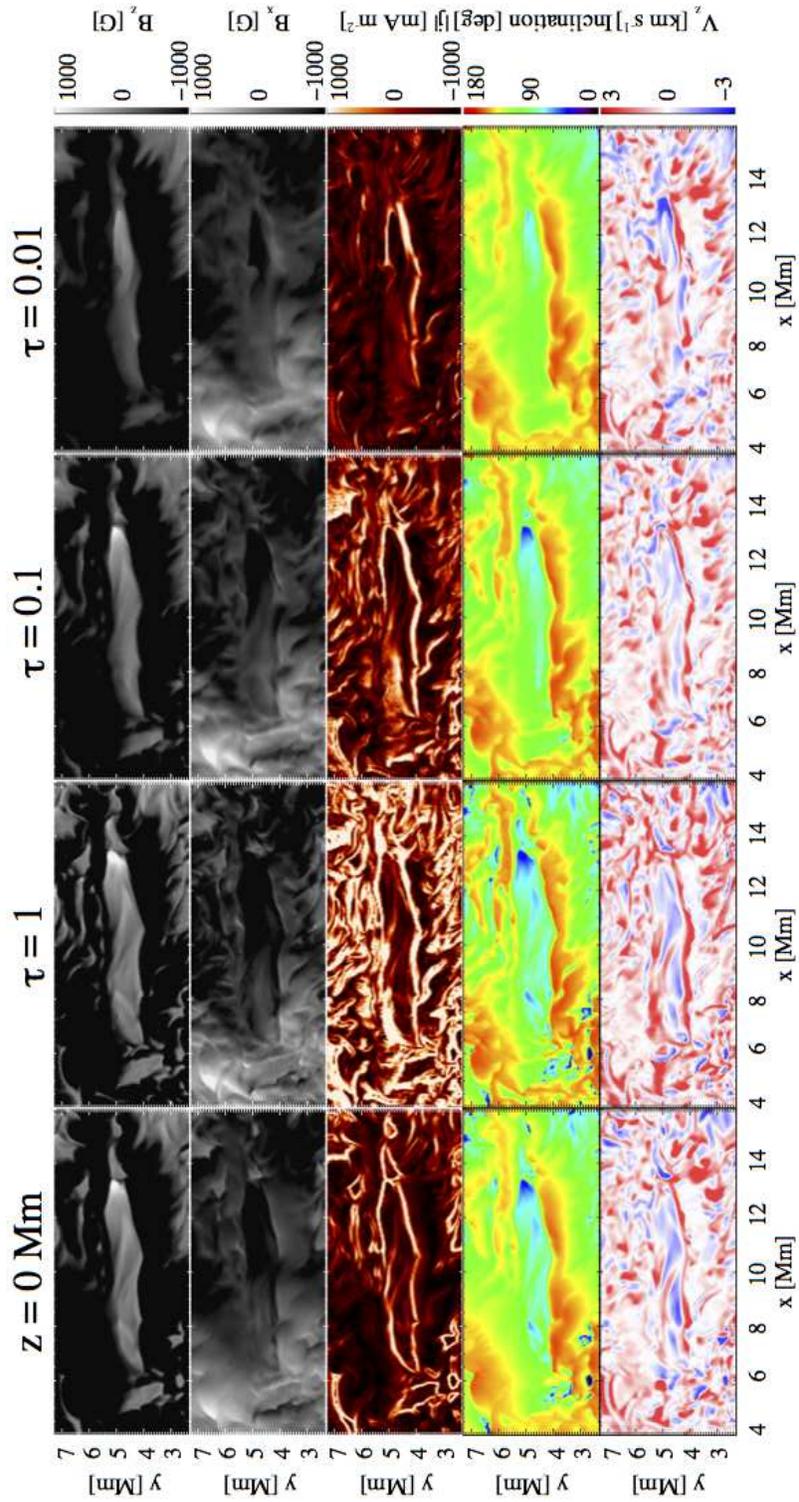}
  \end{center}
  \caption{
    Physical parameters
    measured at a constant geometrical height
    ($z=0\ {\rm Mm}$)
    and three different iso-$\tau$ levels
    ($\tau=1$, $0.1$, and $0.01$).
  }
  \label{fig:lb_taudepend}
\end{figure}

In Section \ref{subsec:structure},
we calculated the vertical current $|j_{z}|$
using horizontal derivatives
of magnetic fields
measured at a constant geometrical height
($z=0\ {\rm Mm}$).
However,
in the actual observations,
magnetic fields
are measured
at constant-$\tau$ layers,
which may fluctuate vertically.
Therefore,
the observational $|j_{z}|$
may be affected
by such vertical fluctuations.
In fact,
as we saw in Figure \ref{fig:lb_slice},
the iso-$\tau$ levels
are lifted up substantially
inside the light bridge
because of the dense convective upflow.

\begin{figure}
  \begin{center}
    \includegraphics[width=150mm]{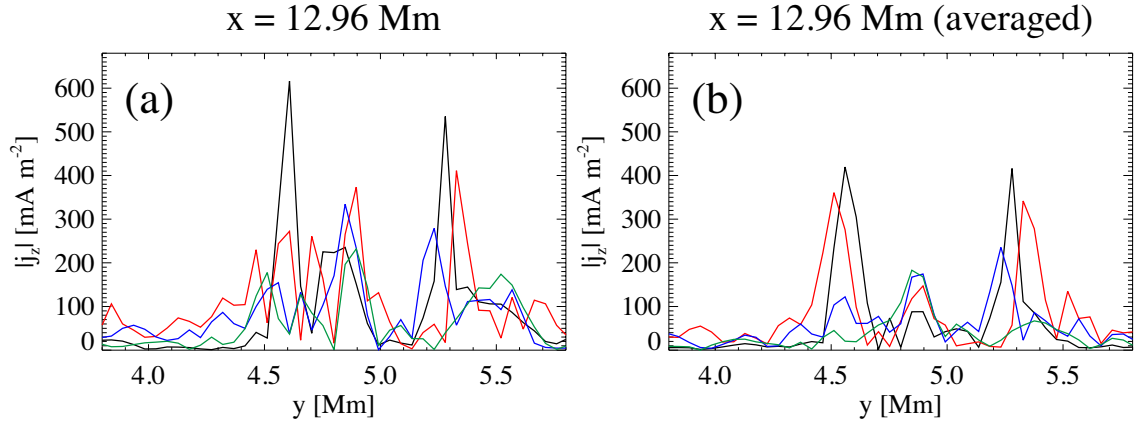}
  \end{center}
  \caption{
    (a) Vertical current $|j_{z}|$
    along $y$-axis (at $x=12.96\ {\rm Mm}$)
    derived from the horizontal magnetic fields
    measured at $z=0\ {\rm Mm}$
    (black: corresponding to purple dashed line
    in Figure \ref{fig:lb_1d}),
    $\tau=1$ (red),
    $\tau=0.1$ (blue),
    and $\tau=0.01$ (green).
    (b) Same as (a) but averaged
    from $x=12.72\ {\rm Mm}$ to $13.2\ {\rm Mm}$.
  }
  \label{fig:lb_1d_taudepend}
\end{figure}

Figure \ref{fig:lb_1d_taudepend}(a) compares
vertical currents
computed from the horizontal fields
measured at different layers.
Although there are some fluctuations
in the iso-$\tau$ currents,
the peak locations and values
roughly agree with the current
measured at the $z=0\ {\rm Mm}$ level.
Such agreements are clearer
if we take averages
over some range
as in panel (b).
Therefore,
we can say that
the observational $|j_{z}|$,
which is derived from
magnetic fields
at a constant-$\tau$ layer,
is still comparable to
the actual $|j_{z}|$.




\clearpage



\clearpage

\clearpage

\clearpage

\begin{deluxetable}{lcc}
\tabletypesize{\scriptsize}
  \tablecaption{Comparison
    of Photospheric Parameters
    between the Observed and Modeled Light Bridges
    \label{tab:comparison}}
  \tablewidth{0pt}
  \tablehead{
    \colhead{Parameter} &
    \colhead{{\it Hinode} Observation\tablenotemark{a}} &
    \colhead{MHD simulation}
  }
  \startdata
  \sidehead{Light bridge}
  Size & $\sim 22\ {\rm Mm}\times 3\ {\rm Mm}$ &
   $\sim 7.5\ {\rm Mm}\times 1.5\ {\rm Mm}$\\
  Lifetime & $\sim 2\ {\rm days}$ & a few hours\\
  $B_{z}$ & $\sim 0\ {\rm G}$ & $\sim 0\ {\rm G}$\\
  $B_{x}$ & $\sim -1000\ {\rm G}$ & $\sim -1000\ {\rm G}$\\
  Inclination & $\sim 90^{\circ}$ & $\sim 90^{\circ}$\\
  $V_{z}$ & $\gtrsim 1\ {\rm km\ s}^{-1}$ & $\sim 2\ {\rm km\ s}^{-1}$\\
  $V_{x}$ & a few km s$^{-1}$ & $\sim\pm 4\ {\rm km\ s}^{-1}$\\
  \sidehead{Light bridge boundary}
  $V_{z}$ & $\gtrsim -6\ {\rm km\ s}^{-1}$ & $<-3\ {\rm km\ s}^{-1}$\\
  $V_{x}$ & \nodata & $<-8\ {\rm km\ s}^{-1}$\\
  $|\mbox{\boldmath $j$}|$ & \nodata & 500 -- $2500\ {\rm mA\ m}^{-2}$\\
  $|j_{z}|$ & $\gtrsim 100\ {\rm mA\ m}^{-2}$ & $\gtrsim 100\ {\rm mA\ m}^{-2}$\\
  \sidehead{Ambient pore}
  $B_{z}$ & $\sim -2000\ {\rm G}$ & $\lesssim -1500\ {\rm G}$\\
  Inclination & $\sim 180^{\circ}$ & $>120^{\circ}$\\
  $V_{z}$ & $\sim 0\ {\rm km\ s}^{-1}$ & $\sim 0\ {\rm km\ s}^{-1}$\\
  \enddata
  \tablecomments{The heights
    at which the physical parameters
    are measured
    are not necessarily the same
    between the two cases.
    Vertical velocity $V_{z}$
    in the observation
    is Doppler velocity.}
  \tablenotetext{a}{Values are taken
    from \citetalias{tor15a}.}
\end{deluxetable}

\begin{deluxetable}{lccc}
\tabletypesize{\footnotesize}
  \tablecaption{Characteristics
    and Activity Phenomena
    of Umbral Dots,
    Light Bridges,
    and Penumbral Filaments
    \label{tab:magnetoconvection}}
  \tablewidth{0pt}
  \tablehead{
    \colhead{} & \colhead{Umbral dots} &
    \colhead{Light bridges} & \colhead{Penumbral filaments}
  }
  \startdata
  Circumstance & umbra & umbra/inter-pore\tablenotemark{a} & penumbra\\
  Configuration & point-like & elongated & elongated\\
  Flow structure & narrow upflow with & upflow, bi-directional outflow\tablenotemark{b}, & upflow, uni-directional outflow,\\
   & adjacent downflows & and peripheral downflows & and peripheral downflows\\
  Background field & vertical & vertical & inclined\\
  Activity events & \nodata & brightenings/surges & microjets\\
  \enddata
  \tablenotetext{a}{The bridges appear
    in umbrae
    in the fragmenting mature sunspots
    and between merging pores
    in the developing ARs.}
  \tablenotetext{b}{Some authors
    reported the uni-directional flows
    in light bridges
    \citep[e.g.,][]{ber03}.
    The magnetic field structure
    may determine
    the direction
    of the convective flow.}
\end{deluxetable}



\begin{thebibliography}{}
\bibitem[Asai et al.(2001)]{asa01}
  Asai, A., Shimojo, M., Isobe, H., et al.
  2001, \apjl, 562, L103
\bibitem[Beckers \& Schr{\"o}ter(1969)]{bec69}
  Beckers, J.~M. \& Schr{\"o}ter, E.~H. 1969, \solphys, 10, 384
\bibitem[Berger \& Berdyugina(2003)]{ber03}
  Berger, T.~E. \& Berdyugina, S.~V. 2003, \apjl, 589, L117
\bibitem[Biskamp et al.(1995)]{bis95}
  Biskamp, D., Schwarz, E., \& Drake, J.~F.
  1995, Physical Review Letters, 75, 3850
\bibitem[Borrero \& Ichimoto(2011)]{bor11}
  Borrero, J.~M. \& Ichimoto, K.
  2011, Living Reviews in Solar Physics, 8, 4
\bibitem[Borrero et al.(2014)]{bor14}
  Borrero, J.~M., Lites, B.~W., Lagg, A., Rezaei, R., \& Rempel, M.
  2014, \aap, 572, A54
\bibitem[Cheung et al.(2007)]{che07}
  Cheung, M.~C.~M., Sch{\"u}ssler, \& M., Moreno-Insertis, F.
  2007, \aap, 467, 703
\bibitem[Cheung et al.(2008)]{che08}
  Cheung, M.~C.~M., Sch{\"u}ssler, M., Tarbell, T.~D., \& Title, A.~M.
  2008, \apj, 687, 1373
\bibitem[Cheung et al.(2010)]{che10}
  Cheung, M.~C.~M., Rempel, M., Title, A.~M., \& Sch{\"u}ssler, M.
  2010, \apj, 720, 233
\bibitem[Cheung \& Cameron(2012)]{che12}
  Cheung, M.~C.~M. \& Cameron, R.~H. 2012, \apj, 750, 6
\bibitem[Choudhuri(1986)]{cho86}
  Choudhuri, A.~R. 1986, \apj, 302, 809
\bibitem[Clyne et al.(2007)]{cly07}
  Clyne, J., Mininni, P., Norton, A., \& Rast, M.
  2007, New J. Phys., 9, 301
\bibitem[Clyne \& Rast(2005)]{cly05}
  Clyne, J., \& Rast, M. 2005, Proc. SPIE, 5669, 284
\bibitem[Evershed(1909)]{eve09}
  Evershed, J. 1909, \mnras, 69, 454
\bibitem[Fan \& Gibson(2003)]{fan03}
  Fan, Y. \& Gibson, S.~E.
  2003, \apjl, 589L, 105
\bibitem[Hood et al.(2009)]{hoo09}
  Hood, A.~W., Archontis, V., Galsgaard, K., \& Moreno-Insertis, F.
  2009, \aap, 503, 999
\bibitem[Jur{\v c}{\'a}k et al.(2006)]{jur06}
  Jur{\v c}{\'a}k, J., Mart{\'{\i}}nez Pillet, V., \& Sobotka, M.
  2006, \aap, 453, 1079
\bibitem[Katsukawa et al.(2007a)]{kat07a}
  Katsukawa, Y., Yokoyama, T., Berger, T.~E., et al.
  2007, \pasj, 59S, 577
\bibitem[Katsukawa et al.(2007b)]{kat07b}
  Katsukawa, Y., Berger, T.~E., Ichimoto, K., et al.
  2007, Science, 318, 1594
\bibitem[Kosugi et al.(2007)]{kos07}
  Kosugi, T., Matsuzaki, K., Sakao, T., et al.
  2007, \solphys, 243, 3
\bibitem[Lagg et al.(2014)]{lag14}
  Lagg, A., Solanki, S.~K., van Noort, M., \& Danilovic, S.
  2014, \aap, 568, A60
\bibitem[Leake et al.(2014)]{lea14}
  Leake, J.~E., DeVore, C.~R., Thayer, J.~P., et al.
  2014, \ssr, 184, 107
\bibitem[Leka(1997)]{lek97} Leka, K.~D. 1997, \apj, 484, 900
\bibitem[Lites et al.(2013)]{lit13}
  Lites, B.~W., Akin, D.~L., Card, G., et al.
  2013, \solphys, 283, 579
\bibitem[Lites et al.(1991)]{lit91}
  Lites, B.~W., Bida, T.~A., Johannesson, A., \& Scharmer, G.~B.
  1991, \apj, 373, 683
\bibitem[Louis et al.(2008)]{lou08}
  Louis, R.~E., Bayanna, A.~R., Mathew, S.~K., \& Venkatakrishnan, P.
  2008, \solphys, 252, 43
\bibitem[Martinez-Sykora et al.(2015)]{mar15}
  Martinez-Sykora, J., De Pontieu, B., Hansteen, V.~H., \& Carlsson, M.
  2015 (in prep)
\bibitem[Nishizuka et al.(2012)]{nis12}
  Nishizuka, N., Hayashi, Y., Tanabe, H., et al.
  2012, \apj, 756, 152
\bibitem[Parker(1979)]{par79}
  Parker, E.~N. 1979, \apj, 234, 333
\bibitem[Rempel \& Cheung(2014)]{rem14}
  Rempel, M. \& Cheung, M.~C.~M.
  2014, \apj, 785, 90
\bibitem[Rempel et al.(2009a)]{rem09a}
  Rempel, M., Sch{\"u}ssler, M., Cameron, R.~H., \& Kn{\"o}lker, M.
  2009, Science, 325, 171
\bibitem[Rempel et al.(2009b)]{rem09b}
  Rempel, M., Sch{\"u}ssler, M., \& Kn{\"o}lker, M.
  2009, \apj, 691, 640
\bibitem[Rempel(2011)]{rem11}
  Rempel, M. 2011, \apj, 729, 5
\bibitem[Rempel(2012)]{rem12}
  Rempel, M. 2012, \apj, 750, 62
\bibitem[Rimmele(2008)]{rim08}
  Rimmele, T. 2008, \apj, 672, 684
\bibitem[Roy(1973)]{roy73}
  Roy, J.~R. 1973 \solphys, 28, 95
\bibitem[R{\"u}edi et al.(1995)]{rue95}
  R{\"u}edi, I., Solanki, S.~K., \& Livingston, W.
  1995, \aap, 302, 543
\bibitem[Scharmer et al.(2013)]{sch13}
  Scharmer, G.~B., de la Cruz Rodriguez, J., S{\"u}tterlin, P., \& Henriques, V.~M.~J.
  2013, \aap, 553, A63
\bibitem[Sch{\"u}ssler \& V{\"o}gler(2006)]{sch06}
  Sch{\"u}ssler, M. \& V{\"o}gler, A. 2006, \apjl, 641, L73
\bibitem[Shimizu et al.(2009)]{shi09}
  Shimizu, T., Katsukawa, Y., Kubo, M., et al.
  2009, \apjl, 696, L66
\bibitem[Shimizu(2011)]{shi11}
  Shimizu, T. 2011, \apj, 738, 83
\bibitem[Solanki \& Montavon(1993)]{sol93}
  Solanki, S.~K. \& Montavon, C.~A.~P. 1993, \aap, 275, 283
\bibitem[Spruit \& Scharmer(2006)]{spr06}
  Spruit, H.~C. \& Scharmer, G.~B. 2006, \aap, 447, 343
\bibitem[Strous \& Zwaan(1999)]{str99}
  Strous, L.~H. \& Zwaan, C. 1999, \apj, 527, 435
\bibitem[Thomas \& Weiss(1992)]{tho92}
  Thomas, J.~H. \& Weiss, N.~O. 1992,
  in Sunspots: Theory and Observations,
  ed. J. H. Thomas \& N. O. Weiss
  (NATO ASI, C 375; Dordrecht: Kluwer), 3
\bibitem[Toriumi et al.(2015)]{tor15a}
  Toriumi, S., Katsukawa, Y., \& Cheung, M.~C.~M.
  2015, \apj, in prep
\bibitem[Tsuneta et al.(2008)]{tsu08}
  Tsuneta, S., Ichimoto, K., Katsukawa, Y., et al.
  2008, \solphys, 249, 167
\bibitem[V{\"o}gler et al.(2005)]{voe05}
  V{\"o}gler, A., Shelyag, S., Sch{\"u}ssler, M., et al.
  2005, \aap, 429, 335
\bibitem[Zakharov et al.(2008)]{zak08}
  Zakharov, V., Hirzberger, J., Riethm{\"u}ller, T.~L. Solanki, S.~K., Kobel, P.
  2008, \aap, 488L, 17
\bibitem[Zwaan(1985)]{zwa85}
  Zwaan, C. 1985, \solphys, 100, 397
\end{thebibliography}
\end{document}